\documentclass[prb,showpacs,twocolumn,superscriptaddress]{revtex4}
\usepackage[dvipdfmx]{hyperref}
\usepackage{graphicx}
\usepackage{amsfonts,amsmath,amssymb,bm,amsbsy}
\usepackage{upgreek,color}

\newcommand{\myFig}[5]{ %
\begin{figure}[htb] 
\begin{center} 
\includegraphics[width=#1\columnwidth, height=#2\columnwidth, clip=true, keepaspectratio]{#3}
\caption{#4} \vspace{-0.5cm} \label{#5} 
\end{center} \end{figure}} 
\newcommand{\vcr}[1]{\boldsymbol{\mathrm{#1}}}

\begin{document}
\title{On the Newtonian origin of the spin motive force in ferromagnetic atomic wires}

\author{Maria Stamenova}
\affiliation{School of Physics and CRANN, Trinity College, Dublin 2, Ireland} 
\author{Tchavdar N. Todorov}
\affiliation{School of Mathematics and Physics, Queen's University Belfast, Belfast BT7 INN, UK}
\author{Stefano Sanvito} \email[Contact email address: ]{sanvitos@tcd.ie}
\affiliation{School of Physics and CRANN, Trinity College, Dublin 2, Ireland} 

\begin{abstract}
We demonstrate numerically the existence of a spin-motive force acting on spin-carriers when moving in a time and space dependent internal field. This is the case for electrons in a one-dimensional wire with a precessing domain wall. The effect can be explained solely by adiabatic dynamics and is shown to exist for both classical and quantum systems. 
\end{abstract}


\maketitle

\section{Introduction}

Recently Barnes and Maekawa \cite{barnes} have proposed a generalization of Faraday's law to account for a non-conservative force of spin origin. This arises in systems with time-dependent order parameters as a result of Berry phase (BP) accumulation \cite{berry}. As an example they consider a domain wall (DW), formed in a finite ferromagnetic wire and precessing about a static co-axial external magnetic field. In the {\it adiabatic approximation}, where electron spins remain aligned with the local magnetization, a constant potential shift $\Delta \phi$ is generated between the two ends of the wire. This is directly proportional to the angular frequency of precession of the wall $\omega$,
\begin{equation} \label{phi0}
\Delta \phi=\frac{\hbar}{e} \, \omega \, ,
\end{equation}   
and within the Stoner model it exactly cancels the Zeeman potential. Such a potential, described as a spin-motive force (SMF), has been recognised previously in the context of the Aharonov-Casher \cite{ryu,oh} and Stern's  \cite{stern} effects. These are all manifestations of BP related phenomena, where holonomies arise as a result of a parallel transport of some kind \cite{anandan}. The latter does not need to be a quantum effect, another example being the classical Foucault pendulum. 

Here we demonstrate computationally the result of Eq. (\ref{phi0}) through time-dependent quantum-classical simulations of an atomic wire incorporating a precessing DW. We also present an analytical classical argument for the driving mechanism of the SMF in this system. Our approach has the benefit of being ``Berry-phase-free" in the sense that it does not need to call for a Berry phase argument to explain the SMF and demonstrates the Newtonian nature of the conversion of the magnetic response of electronic spins into an electrostatic voltage drop. This is further illustrated with classical dynamical simulations for a system of classical magnetic dipoles in a rotating magnetic field mimicing the DW.
In addition we show that if one abandons the Stoner model and accounts for a non-spin component of the magnetic moments forming the DW, the cancellation between the SMF and the Zeeman potential is incomplete, leaving behind a non-zero net SMF, which can be experimentally measured.  

\section{Model}

We consider a one-dimensional magnetic atomic wire in a magnetic field and describe the conduction electrons by an \textit{s-d} \cite{yosida} tight-binding Hamiltonian
\begin{equation} \label{He}
\hat{\cal H}_e=\sum_{i,j,\alpha}  H_{ij}^{\mathrm {TB}} c_i^{\alpha\dagger}c_j^{\alpha} - \sum_{i,\alpha,\beta} c_i^{\alpha\dagger} {\vcr{\upsigma}}_{\alpha\beta}\, c_i^{\beta} \cdot  \vcr{\Upphi}_{i} \, ,
\end{equation}
where  $c_i^{\alpha\dagger}$ ($c_i^{\alpha}$) is the creation (annihilation) operator for an electron with spin $\pm 1/2$ ($\alpha=1,2$) on the atomic site $i$ and ${\vcr{\upsigma}}$ is the vector of Pauli matrices. The first term in Eq. (\ref{He}) is the spin-independent tight-binding (TB) part, while the second describes the spin interaction with the effective local field ${\vcr{\Upphi}}_{i}$,     
\begin{eqnarray} \label{HTB}
H_{ij}^{\mathrm {TB}} \!\!\!&=& \!\!\! \left( E_0 +  \sum_n \frac{\kappa\Delta q_n}{\sqrt{R_{in}^2+(\kappa/U)^2}} \right) \delta_{ij} + \chi \delta_{i,j\pm 1} \\ \label{Beff}
{\vcr{\Upphi}}_{i} \!\!\!&=& \!\!\! J {\vcr{S}}_i + g_e \mu_B {\vcr{B}} \, , 
\end{eqnarray}
where $E_0$ is the onsite energy ($E_0=0\,\mathrm{eV}$ for all sites), $\kappa=e^2/4\pi\varepsilon_0 = 14.4\,\mathrm{eV\AA}$, $\chi$ is the hopping parameter, $g_e$ is the electron $g$-factor and $\vcr{B}$ is the external magnetic field. The second term in the brackets in Eq. (\ref{HTB}) is a mean-field repulsive electrostatic potential \cite{cristian} with an onsite strength $U$ and a Coulombic decay at large intersite distances $R_{ij}$. $\Delta q_i = q_i - q_i^{(0)}$ is the excess number of electrons on site $i$, $q_i^{(0)}$ being the equilibrium one.

In Eq. (\ref{Beff}) ${\vcr{S}}_i$ is the effective local angular momentum at site $i$, associated with the magnetic moment of that atom, normalized by $\hbar$ (and thus dimensionless). ${\vcr{S}}_i$ are treated as classical variables, nonetheless exchange-coupled with strength $J>0$ to the conduction electrons according to a classical Hamiltonian
\begin{equation}
{\cal H}_\mathrm{S} = -\sum_i \, {\vcr{S}}_i \, \cdot \left[ J \langle {\vcr{s}} \rangle_i + g_{\mathrm S} \mu_B {\vcr{B}} \right] - J_z\sum_i \left( {\vcr{S}}_i \cdot {\hat{\vcr{ z}}} \right)^2 \, .
\end{equation}
Here $g_{\mathrm S}$ is the $g$-factor of ${\vcr{S}}_i$ which could be of mixed spin and orbital origin, $\hat{\vcr{z}}$ ($|\hat{\vcr{z}}|=1$) is the unit vector and $J_z$ the anisotropy constant along the easy $z$-axis. $\langle \vcr{s} \rangle_i= \mathrm{Tr}[\hat{\rho}_{ii} {\vcr{\sigma}}]$ is the expectation value of the electron spin at site $i$, where $\hat{\rho}$ is the density matrix and the trace is over the spin coordinates.

The corresponding quantum and classical Liouville equations of motion for the two subsystems are  
\begin{equation} \label{eom}
\frac{\mathrm{d}\hat{\rho}}{\mathrm{d}t}=\frac{i}{\hbar} \left[ \hat{\rho}, \hat{\cal H}_e \right], \quad \frac{{\mathrm d}{\vcr{S}}_i}{\mathrm{d}t} = \left\{ {\vcr{S}}_i, {\cal H}_S \right\} \, ,
\end{equation}
where $\{\cdot,\cdot\}$ represents the classical Poisson bracket.
\myFig{1.0}{0.6}{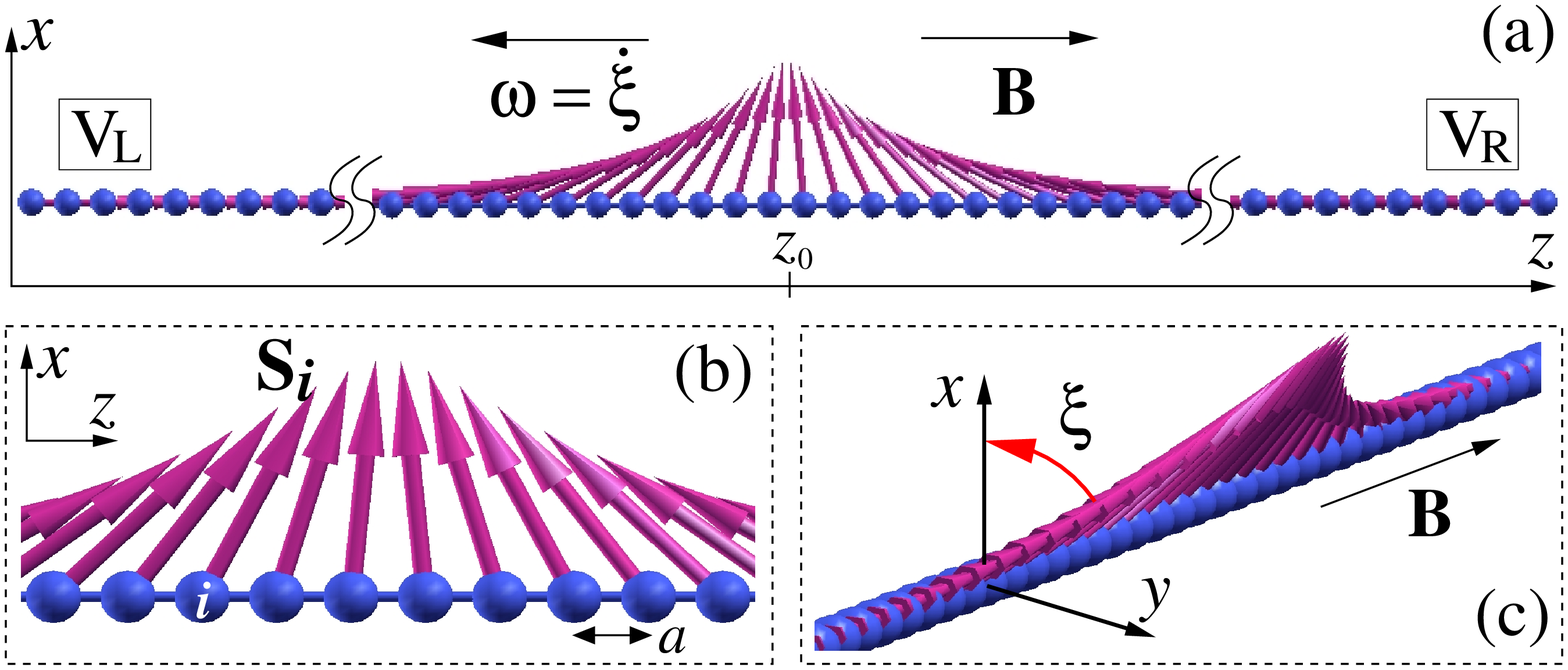}{(Color online) Different prospect views of the DW formed by the local spin $\{\vcr{S}_i\}$ in the middle of a single-atom chain. An external magnetic field, applied along the wire, induces a clockwise rotation of the DW about the $z$-axis.}{system}
These have been solved numerically with a quantum-classical dynamical simulation following the evolution of a preset DW structure in a metallic atomic wire in magnetic field. Initially the set of classical spins $\{\vcr{S}_i\}^{(0)}$ is prepared in a DW arrangement (see Fig. \ref{system}) and relaxed self-consistently in the electronic environment. At time $t=0$ the external magnetic field $\vcr{B}=B {\hat{\vcr{z}}}$ is switched on along the wire and a new initial electronic state is self-consistently determined for $\{\vcr{S}_i\}^{(0)}$. The system is then propagated according to Eqs. (\ref{eom}).

\section{Numerical simulations}

The electrostatic potentials $V_\mathrm{L(R)}$, developing away from the DW on the left(right)-hand-side of the chain, are computed as the spatial (over two identical sets $\mathbb{L}$($\mathbb{R}$) of $N_V$ atoms at each wire end) and temporal (over the evolution time $T$) averages of the onsite potential, i.e. $V_\mathrm{L(R)} (T) = 1/\left( T N_V \right)  \sum\limits_{i \in \mathbb{L(R)}} \int\limits_{0}^{T} \mathrm{d} t \sum\limits_{n=1}^N \frac{\kappa\Delta q_n(t)}{\sqrt{R_{in}^2+(\kappa/U)^2}}$, where $N$ is the total number of atoms. We investigate the stationary voltage drop ${\Delta V}_\mathrm{calc} = \lim_{T\rightarrow \infty} \left[ V_{\mathrm L} (T) - V_{\mathrm R}(T) \right]$ that builds up across the system. In the limit of local charge neutrality ($U\rightarrow \infty$) this should compensate for any arising spatial energy disturbances in the system. We anticipate two such contributions 
\begin{equation} \label{dV}
\Delta V = \Delta \phi - g_e \mu_{\mathrm B} B/e \,   
\end{equation}  
where the first term is due to the proposed non-conservative SMF from Eq. (\ref{phi0}), while the second is due to the Zeeman split. In order to extract the effect of the SMF itself in the first set of simulations we have set $g_e=0$, so that only the first term in Eq. (\ref{dV}) remains.

The parameters used for the simulations are $\chi=-1$~eV, $q_i^{(0)}=1.75$~e/atom for every $i$, $U=7$~eV, $J=1$~eV, $J_z=0.5$~meV, $g_\mathrm{S}=2$, $a\equiv R_{i,i+1}=2.5$~\AA. $N=400$ atoms so that the chain is much longer than the typical width of the relaxed DW (about 10 atomic spacings). The values of $\chi$, $J$ and $q^{(0)}$ are chosen such as to produce a halfmetallic system with a completely filled spin-up band, which lies about $0.5\,\mathrm{V}$ below the Fermi level.

\myFig{1.0}{}{fig02.eps}{(Color online) Time evolution of some dynamical variables at $B=100\,\mathrm{T}$ and for $g_e=0$: (a) $\vcr{S}_x$ and $\vcr{S}_y$ local spin components at the DW center $z_0$, showing the clockwise rotation of the DW about the $z$-axis. The angular frequency $\omega$ of the DW precession is extracted by fitting $S_x (T)$ to $\cos{(\omega T)}$; (b) longitudinal displacement of the DW center $z_0$; (c) averaged potentials $V_\mathrm{L}$, $V_\mathrm{R}$ and $\Delta V_\mathrm{calc}$ (see text).}{qemf}

\myFig{1.0}{0.6}{fig03.eps}{(Color online) Calculated SMF as a function of the DW precession and dependence of the slope over the Coulomb parameter $U$ for $g_e=0$ and $g_\mathrm{S}=2$. (a) The calculated stationary $\Delta V$ depends linearly on $\omega$ with a slope $\hbar_\mathrm{calc} \approx 0.92\hbar$ for realistic values of the parameters $J$ and $U$; (b) $\hbar_\mathrm{calc}$ tends to saturate at the exact value of $\hbar$ with increasing $U$.}{qemf_U}

The spin-DW in these simulations undergoes a steady-rate clockwise rotation with an angular frequency $\omega$ about the direction of the field [see Fig. \ref{qemf}(a)] and exhibits nearly rigid oscillations about a center $z_0$, slightly displaced to the left [see Fig. \ref{qemf}(b)]. The steady rotation generates a SMF, with gentle oscillations that correlate with those of the DW (since the projection of the total spin in the system on the direction of the field is conserved) and which has an asymptotic time-averaged value $\Delta V_\mathrm{calc}$ [see Fig. \ref{qemf}(c)]. The dependence $\Delta V_\mathrm{calc} (\omega)$, obtained by sweeping the external field between 20\,T and 500\,T,  is linear [see Fig. \ref{qemf_U}(a)] with a slope $\hbar_\mathrm{calc} = 0.606\,\mathrm{eV\,fs} \approx 0.92\,\hbar$. The deviation of $\hbar_\mathrm{calc}$ from the exact $\hbar$ [from Eq. (\ref{phi0})] is studied with respect to the two main assumptions in our model: (i) the adiabaticity, which is governed by the strength of the exchange coupling $J$, and (ii) the local charge neutrality, which allows us to identify $\vert\Delta V\vert$ with the SMF and is exact only for $U\rightarrow \infty$. The first criterion is found to be well satisfied for $J=1$~eV. Increasing $J$ ten times results in less than 1\% improvement in $\hbar_\mathrm{calc}$. The ratio $\hbar_\mathrm{calc}/\hbar$, however, is found to be sensitive to $U$ and it asymptotically tends to 1 as $U$ is increased [see Fig. \ref{qemf_U}(b)]. This result confirms the validity of Eq. (\ref{phi0}) and indeed demonstrates that a SMF originates from the precession of the DW.

In reality, however, the effect of the applied magnetic field on the electrons cannot be switched off. We, therefore, return to Eq.~(\ref{dV}) and rewrite it in the form 
\begin{equation} \label{q_emf}
\Delta V = \frac{\hbar}{e} \omega - \frac{g_e}{g_\mathrm{S}} \frac{\hbar}{e} \omega_{\mathrm S} = \left( 1 - \frac{g_e}{g_\mathrm{S}^*} \right) \frac{\hbar}{e} \omega.  
\end{equation} 
Here $\omega_\mathrm{S}=g_\mathrm{S} \mu_B B /\hbar $ is the Larmor frequency of the local spins. The actual angular frequency of precession of the DW $\omega$ differs slightly from $\omega_\mathrm{S}$ due to the exchange interaction with the conduction electrons. In order to account for this effect, we have introduced an effective $g_\mathrm{S}^*$ such that $\omega=g_\mathrm{S}^* \mu_B B /\hbar$. We have verified Eq. (\ref{q_emf}) numerically by varying the value of $g_\mathrm{S}$ (see Fig. \ref{qemf_g}). The effective value $g_\mathrm{S}^*$ is determined by the calculated precession frequency of the wall (Fig. \ref{qemf_g}c). Finally we have again obtained a value of $\hbar_\mathrm{calc} \approx 0.92\,\hbar$, identical to the previous finding in the case $g_e=0$ for this choice of exchange parameter and charging strength.

\myFig{1.0}{0.8}{fig04.eps}{(Color online) Computational demonstration of Eq. (\ref{q_emf}) for $g_e=2$ and a set of values $g_\mathrm{S}=0.5 \div 3$. Panel (a) shows the linear dependence of the stationary potential drop $\Delta V_\mathrm{calc}$ on the angular precession frequency $\omega$; (b) is used to determine the effective $g$-factors $g_\mathrm{S}^*$ and they are compared to the input values $g_\mathrm{S}$ in (c). Note that $g_\mathrm{S}^*=g_\mathrm{S}$ for $g_\mathrm{S}=g_e=2$. Panel (d) demonstrates the validity of Eq. (\ref{q_emf}).}{qemf_g}

Apparently, the voltage drop across the system fully disappears when $g_\mathrm{S}=g_\mathrm{S}^*=g_e$, as derived in reference
[\onlinecite{barnes}] for the Stoner model. However, in $s$-$d$ systems, where $g_\mathrm{S}^*$ has a partially orbital origin, this is not the case and the SMF manifests itself as a measurable quantity. This could be used to determine the effective $g$-factor of the localized spins. In particular if the DW precession is blocked, the measured drop would be just equal to the Zeeman split, i.e. a measurement could determine if the wall is precessing or not. In the remaining part of the paper we examine the mechanism for the SMF by a classical analogy.        

\section{Classical portrait}

Instead of quantum electrons as in the first part of the paper we now consider non-interacting classical particles with an intrinsic angular momentum $\vcr{s}$ ($\vert \vcr{s} \vert = s = \hbar/2$) and with the electron mass $m_e$. These are trapped in a one-dimensional box, where an inhomogeneous time-dependent field
\begin{equation}
\vcr{b} (z,t) = b \left( \cos(f t) \sin(\theta_z),\, \sin(f t) \sin(\theta_z),\,  \cos(\theta_z) \right)
\end{equation}
is present and $\theta_z=\theta(z)$ is chosen such as to mimic a continuous DW-like structure \cite{thetaz}, rotating rigidly with an angular frequency $f$. Here $f$ is analogous to $\omega$ from the quantum simulation, though $f>0$ corresponds to an anticlockwise rotation about the longitudinal axis. The classical Hamiltonian of the spin-particles in the field $\vcr{b}$ is analogous to that of the quantum electrons interacting with local spins $\{\vcr{S}_i\}$, 
\begin{equation}
\mathcal{H}_\mathrm{class} = \frac{p^2}{2 m_e}  - \gamma \vcr{s} \cdot \vcr{b}(z,t)\, , 
\end{equation} 
where $\gamma$ is the coupling strength (analogous to $J$) and $p$ is the canonical momentum of the particles. Then
Hamilton's equations of motion \cite{aharonov} are
\begin{equation} \label{FrcTrq}
m_e \ddot{z} = \gamma \vcr{s} \cdot \nabla_z \vcr{b} (z,t), \quad \dot{\vcr{s}} = \gamma \vcr{s} \times \vcr{b}(z,t)
\end{equation} 
where $\nabla_z \equiv (\partial / \partial z )_{\vcr{s},t} $.    

We consider the limit of large $\gamma b$, in which the dynamics of the spin-particle becomes adiabatic, in the sense that $\vcr{s}$ remains closely aligned with $\vcr{b}$ and its precession about $\vcr{b}$ is by far the fastest motion in the system. However, for $\vcr{s}$ to follow $\vcr{b}(z,t)$, there must always be {\it some} residual misalignment \cite{berger} between the two. This is necessary in order to generate those torques, which, when averaged over the quick precession of $\vcr{s}$, enable $\vcr{s}$ to keep up with $\vcr{b}(z,t)$. This small misalignment, marked by the angle $\varphi$ in Fig. \ref{sBz}, is also the origin of the effective Newtonian force on the spin-particle that manifests itself as SMF.

Fig. \ref{sBz} depicts $\vcr{s}$ and $\vcr{b}$ at some instance of the particle's migration. Differentiating the relation between the angles at the bottom vertex of the tetrahedron $\cos{\alpha} = \cos{\varphi} \cos{\theta} - \sin{\varphi} \sin{\theta} \cos{\beta}$ under the condition $s_{\varphi 1} = s \sin{\varphi} \sin{\beta} = \mathrm{const}$ (which corresponds to keeping $\vcr{s}$ and $t$ fixed), in the adiabatic limit $\varphi\rightarrow 0$, we obtain 
\begin{equation} \label{nablas}
\nabla_z \varphi = - \cos{(\beta)} \, \nabla_z \theta\, .
\end{equation} 

\myFig{1.0}{0.5}{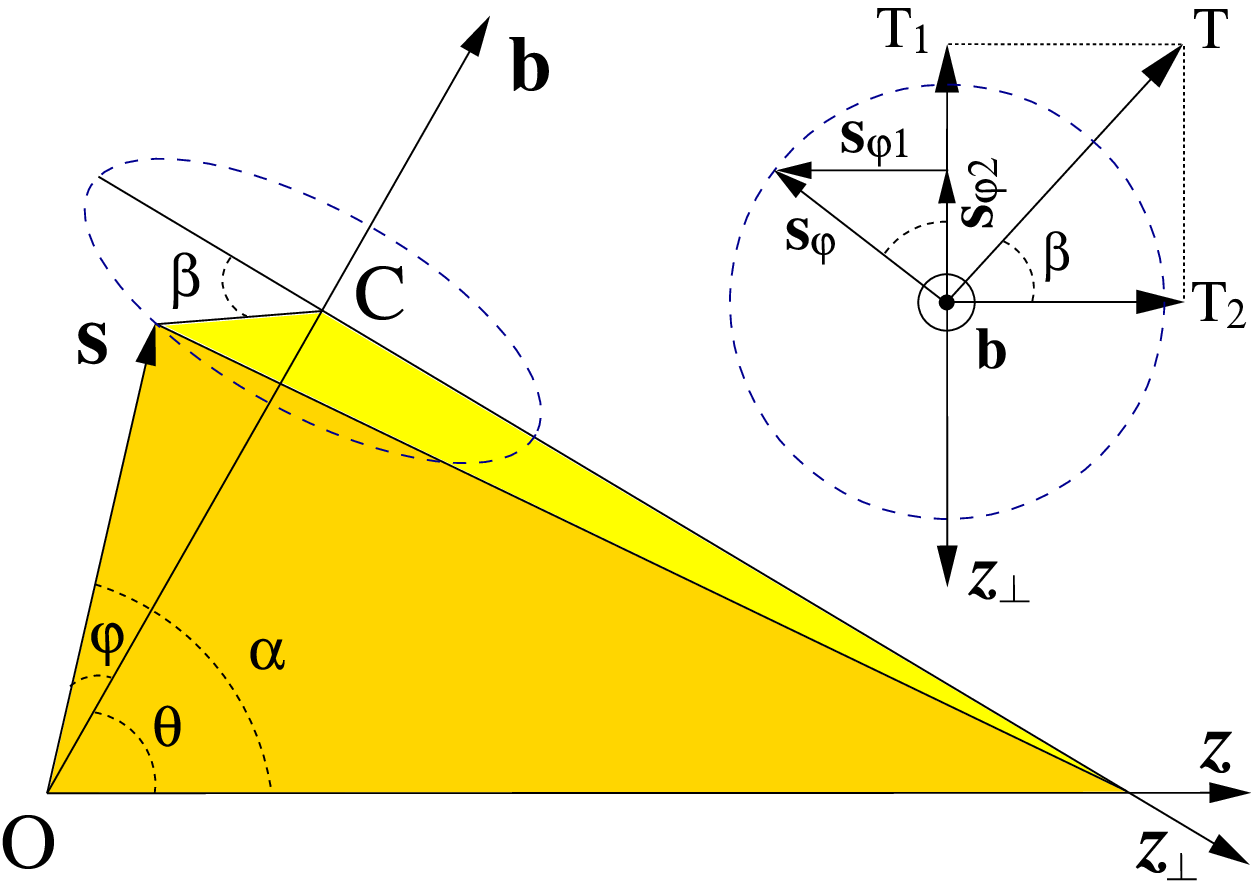}{(Color online) A snapshot of spin-particle's passage through the DW-like region of $\vcr{b}(z,t)$.}{sBz}

From Eq. (\ref{FrcTrq}) the longitudinal force $F_z$ and the torque $T=\vert \vcr{T}\vert=\gamma s_\mathrm{\varphi} b$ are related by  
\begin{equation} \label{frc}
F_z = - \gamma \vert \vcr{s} \vert \vert \vcr{b} \vert \sin(\varphi) \nabla_z \varphi = - T  \nabla_z \varphi = T_2  \nabla_z \theta \, ,
\end{equation} 
where Eq. (\ref{nablas}) has been applied and the full torque $T$ decomposed into two orthogonal torques with magnitudes $T_{1}=T\sin{\beta}$ and $T_{2}=T\cos{\beta}$ (see the inset of Fig. \ref{sBz}).

In the adiabatic regime ($s_{\varphi}\ll s$), we average the two components of the torque over the fast precession of $\vcr{s}$ about $\vcr{b}$. These averaged torques $\overline{T_1}$ and $\overline{T_2}$ must be driving the two separate motions of the spin as the particle crosses the region of the rotating DW-like field, namely, a rotation in the $b$-$z$ plane enabling $\vcr{s}$ to keep up with the spatial variation of $\vcr{b}$, and another rotation in a plane perpendicular to the $z$-axis which makes $\vcr{s}$ follow the anticlockwise precession of $\vcr{b}$, and thus 
\begin{equation} \label{trqs}
\overline{T_1} \approx s \dot{z} \nabla_z \theta, \qquad \overline{T_2} \approx  - \vert \vcr{s} \times \vcr{f} \vert = - s f \sin{\theta} \,.
\end{equation} 

Applying Eq. (\ref{frc}), the averaged linear force upon the spin-particle in the rotating magnetic field is then $\overline{F_z} =\overline{T_2} \nabla_z \theta = s f \sin{(\theta)} \nabla_z \theta $ and therefore the work done by the rotating DW-like field (or the SMF) on the spin-particle for one traversal (left to right) is 
\begin{equation} \label{work}
W_{\mathrm{L} \rightarrow \mathrm{R}} = \int\limits_{z_\mathrm{L}}^{z_\mathrm{R}} \overline{F_z} \mathrm {d} z = - s\, f \int\limits_{0}^{\pi} \sin{(\theta)} \mathrm {d} \theta = - 2 s f \, ,
\end{equation}
where $z_\mathrm{L,R}$ are the leftmost and the rightmost position of the spin-particle on the wire far from the region of spatial variation of the field. This result has been derived with the single assumption of adiabaticity. The adiabatic condition is $s_\varphi \ll s$ and for it to hold it is necessary that the two components of $s_\varphi$ averaged over the rapid precession, simultaneously satisfy the latter, i.e.
\begin{eqnarray} \label{AdC0}
\overline{s_{\varphi 1}} \!&=&\! \overline{T_1}/\gamma b = s v \nabla_z \theta / \gamma b \ll s \:,\\
\overline{s_{\varphi 2}} \!&=&\! \overline{T_2}/\gamma b = s f \cos{(\theta)} / \gamma b \ll s \, . \nonumber
\end{eqnarray}
Thus, considering the maximum attainable values of the right-hand sides and since $\max {(\nabla_z \theta)} = 1/z_\mathrm{w} $, the necessary conditions for adiabaticity are
\begin{eqnarray} \label{AdC}
1/t_\mathrm{w} \ll f_\mathrm{L}, \quad f \ll f_\mathrm{L}
\end{eqnarray} 
where $t_\mathrm{w}$ is the time it takes for the spin-particle to cross the DW-like region of width $z_\mathrm{w}$ \cite{thetaz} and $f_\mathrm{L}=\gamma b$ is its Larmor precession frequency about the field $\vcr{b}$.

\myFig{0.9}{}{fig06.eps}{(Color online) The spatial particle imbalance $\Delta N /N$ and the potential energy difference $\Delta W$ as a function of the angular frequency $f$ of rotation of the DW-like field. The insets represent the change of the velocity and the longitudinal component of the spin of a particle for one left-to-right (red solid) and one right-to-left (blue dashed line) traversal of the wire for $f=0.2\, \mathrm{fs}^{-1}$.} {part_imb}

In order to mimic the typical strength of the exchange interactions ($\sim 1\,\mathrm{eV}$) in our classical simulations, we have used $\gamma=2 \mu_B/\hbar$ and $b=10^5\, \mathrm{T}$. We have simulated an ensemble of $N=700$ noninteracting spin-particles, confined in a $400\,\mathrm{\AA}$-long atomic wire. 
The particles start at random positions within two regions near both wire ends and with velocities identical in magnitude ($v_0=8\,\mathrm{\AA/fs}$)
but random in sign. The DW-like region has $z_\mathrm{w}=5\,\mathrm{\AA}$ (see \cite{thetaz}) which is similar to the $z_\mathrm{w}$, fitted to the relaxed DW profile in the quantum simulation (in atomic spacings) and the typical passage time is $t_\mathrm{w}\approx 1\div 2\,\mathrm{fs}$.  We have used frequencies $\vert f\vert \leq 0.2\,\mathrm{fs}^{-1}$. Since our $f_\mathrm{L}=\gamma b = 17.6\,\mathrm{fs}^{-1}$, these parameters well satisfy the adiabatic conditions of Eq. (\ref{AdC}).

We have integrated numerically Eqs. (\ref{FrcTrq}) and found a stationary difference in the number of particles to the left and to the right of the DW-like region, $\Delta N_\mathrm{calc} = N_\mathrm{R} - N_\mathrm{L}$, developing in time and depending linearly on the frequency $f$ of rotation of the field (see Fig. \ref{part_imb}). By energy conservation, $\Delta N_\mathrm{calc}$ converts to a potential energy shift 
\begin{equation} \label{DelW}
\Delta W_\mathrm{calc}=2 m_e v_0^2 \Delta N_\mathrm{calc} /N
\end{equation}
and the latter is a manifestation of the SMF work $- W_{\mathrm{L} \rightarrow \mathrm{R}}$, derived in Eq. (\ref{work}). Eq. (\ref{DelW}), relating of particle imbalance to SMF, is only valid if the particles have enough initial kinetic energy to traverse the wall from both sides, which,
from one of the sides, means climbing the SMF ramp. Thus the requirement
\begin{equation} \label{KEcond}
m_e v_0^2 /2 > 2 s f
\end{equation} 
sets a lower limit on the initial velocity of the spin-particles in our simulations, for a given $f$. 

Within the adiabatic regime the dependence of $\Delta W_\mathrm{calc}$ on $f$ is found to be linear with a slope of $(0.643\pm 0.012)\,\mathrm{eV\,fs}$ (see the right-hand side scale of Fig. \ref{part_imb}) and agrees with the analytical prediction of  $2s=\hbar=0.658\,\mathrm{eV\,fs}$ in Eq. (\ref{work}).

The directions of the SMF observed in the quantum and the classical simulations agree with the one set by Eq. (\ref{work}), i.e. the SMF is opposite to the direction of the angular velocity of the DW rotation if the itinerant spins are aligned parallel to the local field. Note that with the choice of the band structure in our quantum simulations the effect is carried by the down-spin, so that for all other parameters being equal the sign of the SMF is opposite to that of the classical model. In general, the direction and magnitude of the voltage drop is found to scale with the Fermi level spin-polarization $\eta=(D_{\uparrow}-D_{\downarrow})/ (D_{\uparrow}+D_{\downarrow})$, being $D_{\uparrow (\downarrow)}$ the spin-up(down) density of states at the Fermi level, and $\omega$ as $V_{\mathrm {L} \rightarrow \mathrm{R}} = - \eta \hbar \omega$/e, where $\omega>0$ corresponds to an anticlockwise rotation of the DW (spin) about the $z$-axis. For the half-metallic case studied here $\eta=-1$.

\myFig{0.9}{}{fig07.eps}{(Color online) Plots of the SMF as a function of the DW rotation frequency in units of the Larmor precession frequency of the itinerant spins about the local field. We present the case of dynamics away from the adiabatic regime for both the quantum [panel (a)] and the classical [panel (b)] simulations. Dashed green lines in both panels correspond to a slope of 1.} {nonadiab}

A further similarity between the quantum and the classical simulations, pointing to the classical origin of the ``quantum'' SMF, is that the quantum effect relies strongly on the adiabatic conditions set by Eq. (\ref{AdC}). As illustrated by Fig. \ref{nonadiab}(a) the effect dies out completely above the Larmor precession frequency $\omega_\mathrm{L}=J/\hbar$ of the exchange coupled spins for any choice of band filling. The threshold in the classical case below $f_\mathrm{L}$ [see Fig. \ref{nonadiab}(b)] is an artefact of the classical model and occurs at $f=f_c=(m_e v_0^2/2)/\hbar$ as determined by Eq. (\ref{KEcond}).  

\section{Conclusions}

In conclusion we have demonstrated computationally the presence of a spin-motive force in a quantum-classical system with a spatially and temporally dependent order parameter. Our SMF has the same magnitude as the one described by Barnes and Maekawa \cite{barnes}. We have considered the more general case of an order parameter of mixed spin and orbital character in which case a measurable voltage drop across the system could indicate the presence of the SMF.  We have also presented an analytical classical argument for the mechanism of the SMF in the adiabatic regime. The latter is supported by purely classical simulations of particles with intrinsic angular momentum in a magnetic field with the spatial and temporal properties of the order parameter in the quantum case. The result is the same in magnitude SMF $\Delta \phi = \hbar \omega/e$, where the $\hbar$ factor comes from the magnitude $s$ of the intrinsic angular momenta, considered to represent the electron spin, i.e. $2 s= \hbar$.

\begin{acknowledgements}
This work is sponsored by the Science Foundation of Ireland (grant SFI02/1N7/I175). Authors wish to acknowledge ICHEC and TCHPC for the provision of computational facilities and support.                
\end{acknowledgements}


\begin{thebibliography}{hbp}
\bibitem{thetaz} Typical shape \cite{barnes} of a one-dimensional DW of Heisenberg spins is $\theta_z \equiv \theta(z)=2 \arctan \left(\exp( (z-z_0)/z_\mathrm{w}) \right)$, where $z_0$ is the position of the wall and $z_\mathrm{w}$ is its width.
\end{thebibliography}
\end{document}